\documentclass[prl,superscriptaddress,showpacs,longbibliography,reprint]{revtex4-1}
\usepackage[utf8]{inputenc}
\usepackage{amsmath}
\usepackage{amsfonts}
\usepackage{amssymb}
\usepackage{braket}
\usepackage{dsfont}

\newcommand{\N}{\mathcal{N}}

\newcommand{\G}{\mathcal{G}}

\newcommand{\SC}{S_{2}^{\text{cl}}}

\newcommand{\rhocl}{\rho_{t}^{\text{cl}}}
\newcommand{\qe}{q_{\text{ent}}}
\newcommand{\qc}{q_{\text{coh}}}

\newcommand{\ketbra}[2]{\ket{#1}\hspace{-2.1pt}\bra{#2}}

\DeclareMathOperator{\Tr}{Tr}

\usepackage{graphicx}

\usepackage{color,soul}

\begin{document}
\title{Quantum and classical temporal correlations in $(1 + 1)D$ Quantum Cellular Automata}

\author{Edward Gillman}
\affiliation{School of Physics and Astronomy, University of Nottingham, Nottingham, NG7 2RD, UK}
\affiliation{Centre for the Mathematics and Theoretical Physics of Quantum Non-Equilibrium Systems,
University of Nottingham, Nottingham, NG7 2RD, UK}

\author{Federico Carollo}
\affiliation{Institut f\"{u}r Theoretische Physik, Universit\"{a}t T\"{u}bingen, Auf der Morgenstelle 14, 72076 T\"{u}bingen, Germany}

\author{Igor Lesanovsky}
\affiliation{School of Physics and Astronomy, University of Nottingham, Nottingham, NG7 2RD, UK}
\affiliation{Centre for the Mathematics and Theoretical Physics of Quantum Non-Equilibrium Systems,
University of Nottingham, Nottingham, NG7 2RD, UK}
\affiliation{Institut f\"{u}r Theoretische Physik, Universit\"{a}t T\"{u}bingen, Auf der Morgenstelle 14, 72076 T\"{u}bingen, Germany}

\begin{abstract}
We employ (1 + 1)-dimensional quantum cellular automata to study the evolution of entanglement and coherence near criticality in quantum systems that display non-equilibrium steady-state phase transitions. This construction permits direct access to the entire space-time structure of the underlying non-equilibrium dynamics. It contains the full ensemble of classical trajectories and also allows for the analysis of unconventional correlations, such as entanglement in the time direction between the ``present" and the ``past". Close to criticality, the dynamics of these correlations --- which we quantify through the second-order Renyi entropy --- displays power-law behavior on its approach to stationarity. 
Our analysis is based on quantum generalizations of classical non-equilibrium systems: the Domany-Kinzel cellular automaton and the Bagnoli-Boccara-Rechtman model, for which we provide estimates for the critical exponents related to the classical and quantum components of the entropy. Our study shows that (1 + 1)-dimensional quantum cellular automata permit an intriguing perspective on the nature of classical and quantum correlations in out-of-equilibrium systems.
\end{abstract}

\maketitle

\textbf{Introduction.} Cellular automata (CA) are paradigmatic models for the study of non-equilibrium processes and phase transitions that fall outside the realm of equilibrium statistical mechanics \cite{Grassberger1979,Odor2003,Lubeck2005,Henkel2008}. They also serve as models for complex dynamical processes or even for computation \cite{Wolfram1983,Wolfram2002}. Since their inception, considerable activity has been dedicated to generalizing the CA concept into the quantum domain \cite{Wiesner2009,Cirac2017,Arrighi2019,Hillberry2020} (see e.g. Ref.~\cite{Farrelly2020} for a review). These efforts have resulted in a host of different settings which combine CA dynamics with quantum coherent evolution. Here, we follow a route introduced in Refs.~\cite{Lesanovsky2019,Gillman2020}, which establishes a direct connection between classical probabilistic CA and the discrete-time dynamics of a quantum system on a (1+1)-dimensional lattice, i.e. $(1+1)D$ quantum cellular automata (QCA). The underlying idea is that, for certain models, the dynamics of classical (diagonal) observables can be described  through the Master equation of an associated classical non-equilibrium process. This establishes a well-defined classical limit, which serves as a convenient starting point for investigations into the impact of quantum correlations on non-equilibrium processes in many-body systems \cite{Buchhold2017,Lesanovsky2019,Carollo2019,Gillman2019,Jo2019,Jo2020,Jo2021}. A further appealing aspect is that these systems are realizable on current quantum simulation platforms, such as two-dimensional Rydberg lattice gases \cite{zeiher2016,kim2018,browaeys2020,ebadi2020}.

\begin{figure}[t]
\centering
\includegraphics[width=1\linewidth]{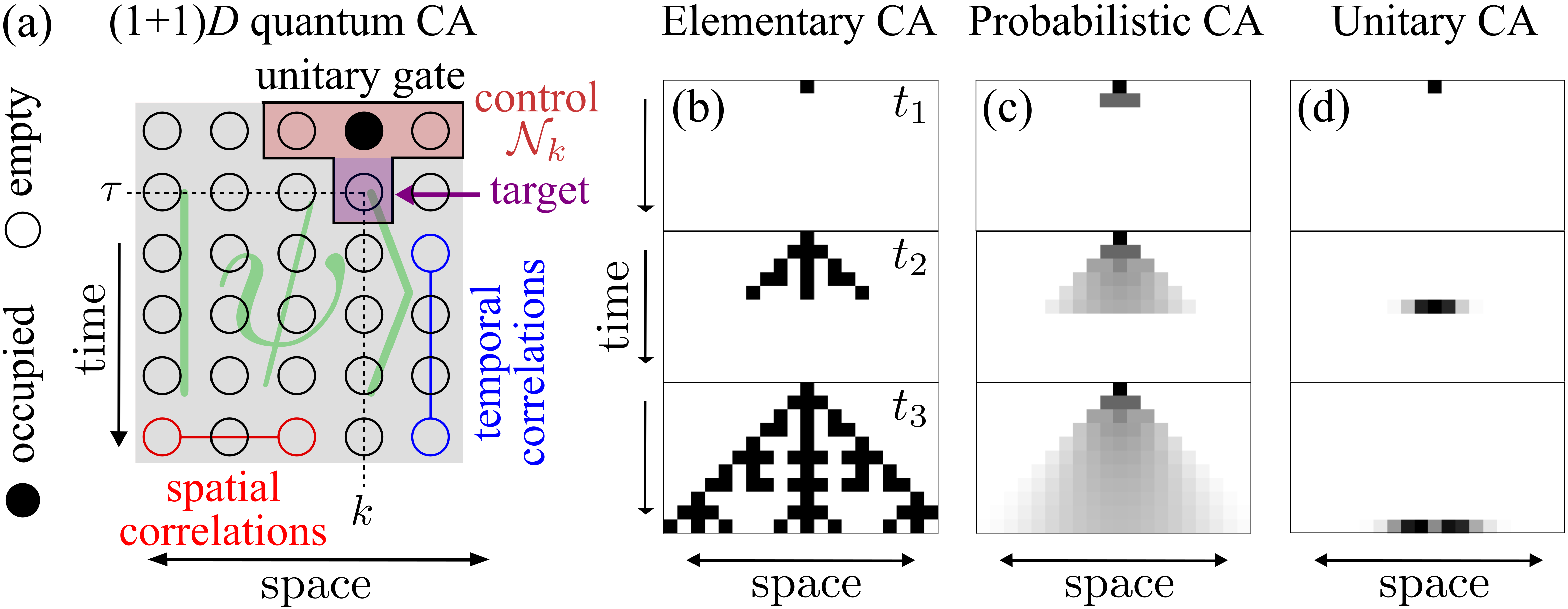}
\caption{\textbf{$\mathbf{(1+1)D}$ quantum cellular automata.} \textbf{(a)} $(1+1)D$ QCA are based on the successive application of unitary quantum gates. By acting on pairs of rows sequentially, an entangled $2D$ quantum state $\ket{\psi}$ is created that encodes the entire space-time structure of the ensuing non-equilibrium dynamics. \textbf{(b)} An elementary cellular automaton (ECA) generates a $2D$ product state. On the QCA an ECA can be realized by gates that merely flip the unoccupied target sites into occupied ones. The image shows snapshots taken at three times ($t_1<t_2<t_3$) for an evolution under rule 150, see Table~\ref{table:gate_rules}. \textbf{(c)} An evolution similar to a probabilistic classical CA is implemented when the gate rotates target sites into superposition states. This generates an entangled $2D$ quantum state. The image shows the density of occupied sites of the BBR model. \textbf{(d)} Unitary evolution of a $1D$ system in the $(1+1)D$ QCA framework, for comparison. Here, the space-time structure for this process is inaccessible.}
\label{fig:Seed_QCA_Schematic_a}
\end{figure}

In this work we exploit that the $(1+1)D$ QCA construction encodes an entire space-time dynamics within a single pure quantum state, i.e.~the whole ensemble of possible histories (space-time trajectories) of a non-equilibrium process becomes accessible by measuring observables of the underlying two-dimensional lattice. This encoding allows to access and analyze new quantities, such as quantum entanglement in the time-domain --- between present and past. We demonstrate this for QCA related to quantum generalizations of two paradigmatic non-equilibrium models: the Domany-Kinzel cellular automaton (DKCA) \cite{Domany1984} and the Bagnoli-Boccara-Rechtman (BBR) model \cite{Bagnoli2001}. Both systems possess a critical point associated with an absorbing phase transition (APT). In its vicinity the second-order Renyi entropy --- which quantifies the present-past entanglement --- displays power-law scaling on its approach to stationarity. While such scaling was observed previously in classical systems \cite{Harada2019}, we show that in the associated quantum models the contribution to the entropy stemming from quantum coherences also displays a power-law decay. The value of the associated critical exponent is such that at long times the entropy is dominated by the classical contribution. This yields not only interesting insights into the dynamical behavior of quantum and classical correlations, but in general highlights the potential offered by the QCA platform which grants access to the full space-time structure of quantum non-equilibrium processes.

\begin{table}
  \centering
  \begin{tabular}{|c|c|c|c|c|c|c|c|c|c|}
  \hline
   $\N$ & $\bullet\bullet\bullet$ & $\bullet\bullet\circ$& $\bullet\circ\bullet$ & $\bullet\circ\circ$ & $\circ\bullet\bullet$ & $\circ\bullet\circ$ & $\circ\circ\bullet$ & $\circ\circ\circ$ \\
    \hline
    ECA &$ 1$ & $0$ &$0$ &$1$&$0$&$1$&$1$&$0$\\
    \hline
    DKCA &$p_{2}$ &$p_{1}$&$p_{2}$&$p_{1}$&$p_{1}$&$0$&$p_{1}$&$0$\\
    \hline
    BBR & $1$ &$p_{2}$&$p_{2}$&$p_{1}$&$p_{2}$&$p_{1}$&$p_{1}$&$0$\\
    \hline 
  \end{tabular}
  \caption{\textbf{Local update rules for $\mathbf{(1+1)D}$ QCA.} The gate \eqref{eqn:gate_def} is defined by eight values of the probability $p(\N)$, one for each possible configuration of the neighbourhood $\N$ [see Fig.~\ref{fig:Seed_QCA_Schematic_a}(a)]. The three examples given encode the classical ECA rule 150, the DKCA and BBR model. The totalistic nature of the updates for the DKCA and BBR model results in only two parameters $p_{1}$ and $p_{2}$. See also Fig.~\ref{fig:Seed_QCA_Schematic_a}(b) and (c) for the evolution of an initial seed generated by the ECA and BBR rules.} \label{table:gate_rules}
\end{table}

\textbf{$\mathbf{(1+1)D}$ QCA and link to other CA models.} QCA are lattice systems with sites that may be either occupied or empty, see Fig.~\ref{fig:Seed_QCA_Schematic_a}(a). The dynamics is constructed such that the occupation of sites in a row --- or more generally in a $d$-dimensional surface --- is determined by those of the row above it \cite{Lesanovsky2019,Gillman2020}. This leads to an effective time-dimension, and the corresponding models are termed $(d+1)$-dimensional with $d$ spatial dimensions perpendicular to the single ``time" dimension \cite{Henkel2008}. Similar models have been studied extensively in the context of classical systems \cite{Hinrichsen2000}. In $(1+1)D$ QCA, at any time $t$ the system is described by a pure quantum state $\ket{\psi_{t}}$, which is an element of the Hilbert space $\mathcal{H} = \bigotimes_{\tau,k} \mathcal{H}_{\tau,k}$, where $\mathcal{H}_{\tau,k}$ are local Hilbert spaces of a $2D$ lattice indexed by ($\tau , k$). Here we consider local Hilbert spaces that are two-dimensional with basis $\lbrace \ket{\circ},\ket{\bullet} \rbrace$, where $n\ket{\circ} = 0, ~ n\ket{\bullet} = \ket{\bullet}$ and $n$ is the local particle number operator. We will refer to the states $\ket{\circ}$ and $\ket{\bullet}$ as occupied and empty respectively. 

The $(1+1)D$ QCA evolves under the action of unitary gates $G_{\tau,k}$. These apply local updates to a ``target" site at $(\tau,k)$, depending on the state of a set of ``control" sites in row $\tau-1$ that form the ``neighbourhood", $\N_{k}$, of site $k$, see Fig.~\ref{fig:Seed_QCA_Schematic_a}(a). Since we are considering binary variables, $\N_{k}$ is taken to be an integer labelling the possible neighbourhoods, whose binary representation (see Table \ref{table:gate_rules}) specifies the occupation of the sites \cite{Wolfram2002}. The state $\ket{\psi_{t}}$ evolves in discrete time-steps as $\ket{\psi_{t}} = \G_{t}\ket{\psi_{t-1}}$. Here the ``global update" $\G_{t}$ consists of an ordered product of $G_{t,k}$, one per each site of row $t$. We will consider the situation in which the $2D$ state is initialised at $t=1$ into a product state of all unoccupied sites, $\ket{\circ}$, except for the first row, $\tau = 1$. This will have a single occupied site, $\ket{\bullet}$, at the center, which we refer to as the seed initial condition. During the subsequent evolution, $\ket{\psi_{t}}$ then encodes the entire space-time structure of cellular automaton dynamics from this initial condition, i.e.~it allows for access to the full history of trajectories, permitting the analysis of typically inaccessible (quantum) correlations between different space-time regions.

The $(1+1)D$ QCA framework generalizes classical cellular automata (CCA) into a unitary quantum setting. It includes canonical classical models, such as deterministic CCA --- e.g.~the much-studied elementary cellular automata (ECA) \cite{Wolfram1983,Wolfram2002} --- and classical probabilistic cellular automata (PCA), such as those studied in the context of APTs \cite{Henkel2008}, as limiting cases [see Fig.~\ref{fig:Seed_QCA_Schematic_a}(b,c)]. Moreover, $1D$ unitary CAs, as discussed in e.g.~Refs.~\cite{Farrelly2020, Hillberry2020} can also be represented as $(1+1)D$ QCA. However, as illustrated in Fig.~\ref{fig:Seed_QCA_Schematic_a}(d), such an evolution generates an ``effectively" $1D$ quantum state (see Ref.~\cite{SM} for details). 

In the following, we consider gates that act on a single target site and a three-site ``control" neighbourhood with the form,
\begin{align}
G_{\tau,k}\ket{\circ} &= \sum_{\N_{k}} P_{\N_{k}} \otimes e^{-i \sigma_x \alpha(\N_{k})}\ket{\circ}.
\label{eqn:gate_def}
\end{align}
The symbol $\otimes$ separates the operators which act on the controls (placed to the left) and on the target (placed on the right): the $P_{\N_{k}} = \ketbra{\N_{k}}{\N_{k}}$ are projectors onto the different ``classical" configurations of the control sites [see Fig.~\ref{fig:Seed_QCA_Schematic_a}(a) and Table \ref{table:gate_rules}] and $\sigma_{x} = \ketbra{\circ}{\bullet} + \ketbra{\bullet}{\circ}$ acts on the target site. The angles $\alpha(\N_{k})$ determine by how much the target site is rotated from its initial state $\ket{\circ}$ into a superposition. The probability for the target site to be occupied, given a particular configuration of the control sites, is $P\lbrace n_{\tau,k} = 1 | \N_{k}\rbrace = p(\N_{k}) = \sin^{2}\left[\alpha(\N_{k})\right]$. For the gate \eqref{eqn:gate_def}, a local update (or ``rule") is specified by choosing the eight (real) values of $\alpha(\N_{k})$ or equivalently $p(\N_{k})$. As we discuss in the following this includes several informative cases, listed in Table \ref{table:gate_rules}, which directly connect to CCA with simultaneous updates. This is because for these cases the $G_{\tau,k}$ for different $k$ commute, such that all choices of $\G_{t}$ are equivalent. 

Choosing rotation angles such that $p(\N_{k})=0,1$, the gate of Eq.~\eqref{eqn:gate_def} reproduces deterministic CCA on the $2D$ state $\ket{\psi_t}$, which remains in a product (unentangled) form at all times, see Fig.~\ref{fig:Seed_QCA_Schematic_a}(b). For example, ECA can be realised in this setting including irreversible cases \cite{Wolfram2002}, which corresponds to the classical result that irreversible CCA can be embedded in higher-dimensional reversible CCA \cite{Toffoli1990}. Further to that, Eq.~\eqref{eqn:gate_def}, contains the case of classical PCA [see Fig.~\ref{fig:Seed_QCA_Schematic_a}(c)]. Here the diagonal components of the density matrix $\Xi_{t} = \ketbra{\psi_{t}}{\psi_{t}}$ are equal to the probability of producing the corresponding space-time configuration under a PCA dynamics \cite{Lesanovsky2019}. However,  the unitary dynamics of the $(1+1)D$ QCA can generate off-diagonal terms in $\Xi_{t}$, i.e., coherence. This means that the gate in Eq.~\eqref{eqn:gate_def} generalizes any desired PCA into a genuine quantum system: the $(1+1)D$ QCA encodes the original classical dynamics, including all associated physics such as APTs, while also displaying uniquely quantum features. For example, the BBR model is a PCA with a three-site control neighbourhood that displays APTs \cite{Bagnoli2001,Bagnoli2014}. It displays two absorbing states (the fully occupied and fully empty product states) and is totalistic, meaning that the local updates depend only on the total number of occupied sites in the neighbourhood, see Table \ref{table:gate_rules}. The corresponding totalistic update rule with a two-site neighbourhood, the DKCA, which displays a single absorbing state of all empty sites, can also be considered in this three-site neighbourhood setting, e.g., by  making updates not depending on the control site in the middle. As for deterministic CCA, $\ket{\psi_t}$ encodes also in this case the entire space-time history, although now being an entangled state, see Fig. \ref{fig:Seed_QCA_Schematic_a}(c).

\textbf{Scaling of temporal correlations.} In the following, we will focus on $(1+1)D$ QCA with gates as in Eq.~\eqref{eqn:gate_def}, that realize quantum generalizations of the DKCA and the BBR model. We will refer to these as QDKCA and QBBR model respectively. At any time, the QCA can be partitioned in such a way that the $t$-th row is singled out. Since rows with $\tau>t$ are in a product state, the entanglement between the two subsystems generated by the partition quantifies the amount of quantum correlations in the pure state $\ket{\psi_t}$, between the ``present" (sites in row $t$) and the ``past" (rows $\tau<t$). We measure this ``present-past" entanglement through the second-order Renyi entropy, i.e. the logarithm of the purity of the reduced state $\rho_{t} = \Tr_{\tau \neq t} \Xi_{t}$ of the $t$-th row:
\begin{align}
S_{2}(t) &= - \ln \Tr \left[ \rho_{t}^{2} \right].
\label{ren-2}
\end{align} 
Given that we are considering extensions of PCA, it is natural to expand $\rho_{t}$ into a diagonal part, $\rho_{t}^{\text{cl}}$, and an off-diagonal part, $X_{t}$, such that $\rho_{t} = \rhocl+ X_{t}$. For classical processes all off-diagonal terms are zero and $\rho_{t} = \rhocl$ \cite{Harada2019,SM}. Under this decomposition the purity of the reduced state becomes a sum of two terms: 
$\Tr \left[ \rho_{t}^{2} \right] = \Tr\left[(\rhocl)^{2}\right] + C_{2}\left(\rho_t\right)$. The first one is equivalent to the classical component of the purity. This can be viewed as due to the probabilistic nature of the process through which sites can be rotated into the occupied state $\ket{\bullet}$. The second term 
$$
C_{2}\left(\rho\right) = \Tr\left[X^{2}\right]=\sum_{i,j} |X_{ij}|^{2} ~ ,
$$
is the $\ell_2$-norm of the density matrix coherence. This contribution is positive, and zero only if $\rho_{t} = \rhocl$. As such, it can only increase the purity of $\rho_t$ and is a manifestation of the quantum correlations present in the QCA. While sufficient for our purposes, we note that $C_{2}$ is not a strict measure of coherence as it violates certain monotonicity conditions \cite{Baumgratz2014}.

For some classical PCA, it was recently found that near the critical point of an ATP the second-order Renyi entropy scales as $S_{2}^{\text{cl}} \sim t^{-p}$, with a universal exponent $p = 0.632613(6)$ \cite{Harada2019}. The same behavior is expected, by construction, when considering $\rhocl$ of the QDKCA and the QBBR model, and hence
\begin{align}
S_{2}^{\text{cl}}(t) = - \ln \Tr \left[(\rhocl)^{2}\right] 
\end{align}
should display a power-law decay at criticality. 

In what follows, we focus on the quantum Renyi entropy in Eq.~\eqref{ren-2} and show that, in the vicinity of the APT of the QDKCA and the QBBR model, it obeys a scaling form
\begin{align}
S_{2}(t) \sim t^{-q_{\text{ent}}}.
\end{align}
Furthermore, we observe that the coherence also follows a similar power-law decay close to criticality,
\begin{align}
C_{2}(t) \sim t^{-\qc},
\end{align}
which defines an additional critical exponent $\qc$. For the seed initial conditions considered, we establish that $\qc \gg \qe$. As might then be expected, we find that the quantum entropy $S_{2}(t)$ tends to the one of the corresponding classical PCA, $S_{2}^{\text{cl}}(t)$, for sufficiently long times.

\begin{figure*}[t]
\centering
\includegraphics[width=1\linewidth]{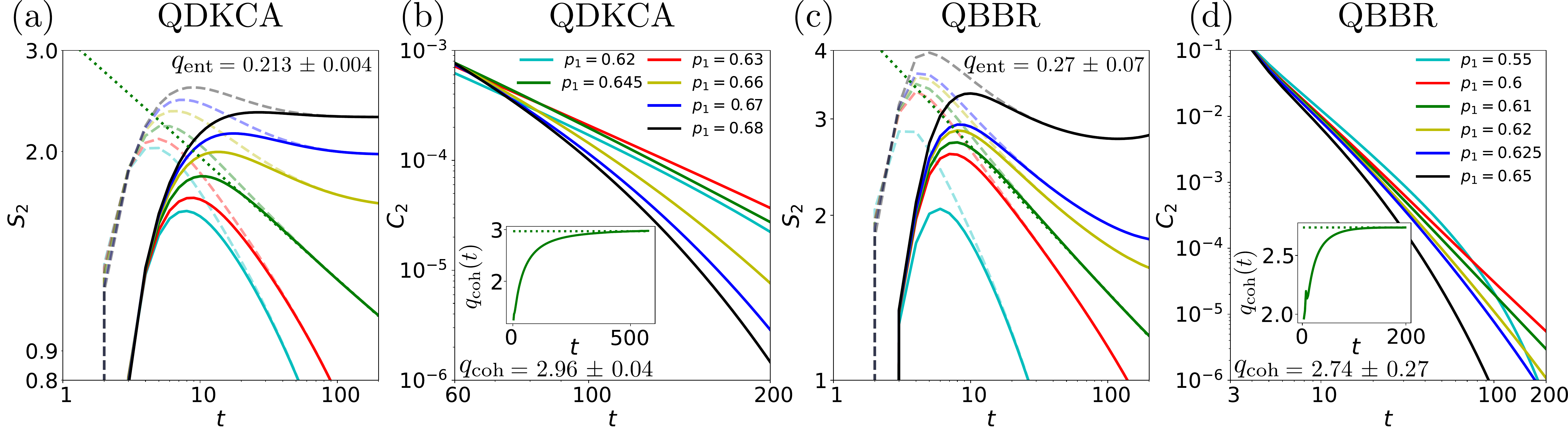}
\caption{\textbf{Critical Scaling of entanglement and coherence:} \textbf{(a)} Dynamics of the second-order Renyi entropy $S_{2}$ (solid lines) for the QDKCA with $p_{2} = 0.874$. MPSs with bond dimension $\chi = 128$ were used to simulate the dynamics of a lattice with $L=256$ sites in the spatial direction. Six different values of $p_{1}$ [indicated in panel (b)] in the vicinity of the critical point were considered. For comparison we also display $\SC$ (dashed lines). For $p_{1} = 0.645$, where the DKCA has a critical point \cite{Henkel2008}, a power-law is observed with exponent $\qe = 0.213 \pm 0.004$. See the text for more details on the estimation procedure and the Supplemental Material for a discussion of error estimation. \textbf{(b)} $C_{2}$ for the QDKCA. At the critical point, $p_{1} = 0.645$, a power-law behaviour is observed. The corresponding exponent is estimated by calculating the effective exponent (see main text), shown in the inset, yielding $\qc = 2.96 \pm 0.04$. \textbf{(c)} $S_{2}$ for the QBBR model and $p_{2} = 0.2$. For $p_{1} = 0.61$, a power-law behaviour with $\qe = 0.27 \pm 0.07$ can be observed. \textbf{(d)} $C_{2}$ for the QBBR model. At the critical point, $p_{1} = 0.61$, power-law scaling with an exponent $\qc = 2.74 \pm 0.27$ is observed. The inset shows the time-dependent effective exponent at criticality.}
\label{fig:Results}
\end{figure*}

\textbf{Numerical Results.} In the following we present results concerning the scaling of $S_{2}$ and $C_{2}$ close to criticality for the QDKCA and the QBBR. We compute these quantities by simulating  the reduced evolution of $\rho_{t}$ using tensor networks (TNs) and matrix product states (MPSs) \cite{Schollwock2011,Montangero2018}. The method employed --- detailed in Ref.~\cite{SM} --- relies on representing $\rho_{t}$ as an MPS and $\G$ as a matrix product operator (MPO), and applying  standard methods for simulating MPS evolution \cite{Paeckel2019,Ran2020}. Details concerning the lattice size $L$, MPS bond-dimension $\chi$ and other parameters related to the simulation are contained in the caption of Fig.~\ref{fig:Results}. All simulations start with an initial seed placed in the center of the first time-slice [see Fig.~\ref{fig:Seed_QCA_Schematic_a}(a)].

In Fig.~\ref{fig:Results}(a) we show the entropy $S_{2}$ for the QDKCA with $p_{2} = 0.874$ and the six values of $p_{1}$ indicated in the figure (solid lines). For the two lowest values of $p_1$, the second-order Renyi entropy rapidly vanishes, due to the fact that the systems approaches an absorbing (product) state. In contrast, the curves with the three highest values of $p_1$ tend to stationary values. This demonstrates that $S_{2}$ can play the role of an order parameter, by distinguishing between the two different phases. When choosing $p_{2} = p_{1}(2-p_{1})$ the DKCA is equivalent to so-called bond-directed-percolation and extensive studies of this classical process have determined the location of the critical point as $p_{2} = 0.874$, $p_{1} = 0.645$ \cite{Henkel2008}. The corresponding critical curve is shown as solid green lines in Fig. \ref{fig:Results}(a) and (b). Here, $S_{2}$ follows a power-law behavior. By fitting the curve between $t = [50,200]$, we estimate the  critical exponent to be $\qe = 0.213 \pm 0.004$. 

Note that, in estimating critical exponents for APTs, determining the location of the critical point is typically a key source of error \cite{Henkel2008}. However, since the QDKCA and DKCA share the same critical point by construction, in our case this error is negligible. The relevant error sources for the QDKCA model are thus associated with finite-$L$, finite-$\chi$ and finite-time effects. In contrast, due to the relatively few studies on the BBR model, for the QBBR model the uncertainty on the location of the critical point provides a significant contribution to the error. Overall, in addition to larger finite-$\chi$ errors, the error associated to the estimates of the critical exponents for the QBBR model are considerably larger than those of the QDKCA. For details on the estimation of errors, and a further discussion of related issues, see Ref.~\cite{SM}.

Fig.~\ref{fig:Results}(a) also displays $S_{2}^{\text{cl}}$ (dashed lines). For each $p_1$, as $t$ increases, the curves for $S_{2}$ and $S_{2}^{\text{cl}}$ become indistinguishable on the scale shown. This means that the critical exponent found also holds for the classical model. We remark that this is different from the value obtained for $S_{2}^{\text{cl}}$ in Ref. \cite{Harada2019}, which may be due to the fact that we are not using a homogeneous initial condition but an initial seed instead. The agreement between $S_{2}(t)$ and $\SC(t)$ suggests that $C_{2}$ becomes irrelevant compared with $\Tr\left[(\rhocl)^{2}\right]$ over time \cite{SM}. This is confirmed in Fig.~\ref{fig:Results}(b), where $C_{2}$ is shown for the same values of $p_{1}$. As with $S_{2}$, at criticality we observe a power-law behaviour in $C_{2}$. However, we find that the timescale over which $C_{2}$ approaches a power-law is considerably larger than that of $S_{2}$. As such, to estimate the critical exponent, we construct the time-dependent effective exponent, $\qc(t) = -\ln\left[C_{2}(t)/C_{2}(t/2)\right]$, as shown in the inset of Fig.~\ref{fig:Results}(b). As can be seen, $C_{2}$ does indeed approach a power-law (indicated by $\qc(t)$ approaching a constant value), and we estimate the exponent to be $\qc = 2.96 \pm 0.04$ by averaging over the effective exponent between $t = [450,550]$. 

Fig.~\ref{fig:Results}(c) and (d) display $S_{2}$ and $C_{2}$ for the QBBR model. In each case we set $p_{2} = 0.2$ and choose six values for $p_{1}$ as indicated in the legend. When $p_{1} = 0.610$ (solid green line) a power-law can be observed, with estimated exponent $\qe = 0.27 \pm 0.07$, obtained by fitting a curve between $t = [50,200]$. Moreover, as $t$ increases, $S_{2}$ and $S_{2}^{\text{cl}}$ become indistinguishable. This is explained by the decay of $C_{2}$, shown in Fig.~\ref{fig:Results}(d). Using the effective exponent for $p_{1} = 0.610$ (shown in the inset), we estimate $\qc = 2.74 \pm 0.27$, by averaging over $t=[150,200]$.  

\textbf{Conclusions and Outlook.} QCA constitute a platform that allows to realize a number of canonical CA scenarios. They can be experimentally realized on quantum simulators and encode the entire space-time information of a non-equilibrium process in a single quantum state. This permits experimental access to unusual properties, such as entanglement in the time domain. Already simple QCA, which are quantum generalizations of the classical DKCA and the BBR model, reveal intriguing features, % such as different scaling behavior of classical and quantum part of the second-order Renyi entropy at criticality.
such as power-law scaling of entanglement and coherence with time at criticality. In the future it would be interesting to focus on more intricate situations, e.g.~QCA where the elementary gates do not commute, so that the order in which local updates are applied defines inequivalent global updates $\G_{t}$ \cite{Gillman2020}. In such a setting, the updates in $(1+1)D$ QCA can be considered as asynchronous updates, the impacts of which have been extensively studied in the classical case \cite{Boure2012,Bandini2012,Fates2013} but are still largely unexplored in the quantum domain. 

\textbf{Acknowledgments.} We acknowledge support from EPSRC [Grant No. EP/R04421X/1], from the ``Wissenschaftler R\"{u}ckkehrprogramm GSO/CZS" of the Carl-Zeiss-Stiftung and the German Scholars Organization e.V., as well as through The Leverhulme Trust [Grant No. RPG-2018-181], and the Deutsche Forschungsgemeinschaft through SPP 1929 (GiRyd), Grant No. 428276754, as well as through Grant No. 435696605. We are grateful for access to the University of Nottingham's Augusta HPC service. 

\bibliographystyle{apsrev4-1}
\bibliography{QCAbib}

% SUPP MATERIAL
\onecolumngrid
\newpage

\pagebreak
\widetext

\begin{center}
\textbf{\large Supplemental Material}
\end{center}

\setcounter{section}{0}
\setcounter{equation}{0}
\setcounter{figure}{0}
\setcounter{table}{0}
\setcounter{page}{1}
\makeatletter

\renewcommand\thesection{S\arabic{section}}
\renewcommand{\theequation}{S\arabic{equation}}
\renewcommand{\thefigure}{S\arabic{figure}}
\renewcommand{\thetable}{S\arabic{table}}
\renewcommand{\bibnumfmt}[1]{[S#1]}

\section{Reduced Evolution of the $1D$ quantum state of row $t$} 
In this section we show how, starting from the dynamics of a $(1+1)D$ QCA, it is possible to obtain the time-evolution of a $1D$ QCA. This is achieved by focusing on the reduced state of the last updated row at every discrete time.

\subsection{Discrete update equation of $1D$ QCA}
For the $2D$ state $\ket{\psi_{t}}$ we can construct the density matrix $\Xi_{t} = \ketbra{\psi_{t}}{\psi_{t}}$, so that the reduced state of row $t$ is readily obtained as  $\rho_{t} = \Tr_{\tau \neq t}\left[\Xi_{t}\right]$, i.e.~by tracing out all the degrees of freedom of the $(1+1)D$ QCA but those in row $t$. Such a reduced state can then be related to the reduced state of the row $t-1$ at time $t-1$, $\rho_{t-1} = \Tr_{\tau \neq t-1}\left[\Xi_{t-1}\right]$, to define a discrete-time evolution equation for the $1D$ state $\rho_{t}$. 

To this end, we note that, from row $t+1$ onward, the state $\ket{\psi_{t}}$ features all sites in the empty state and is thus in a product form. Furthermore, the global update gate $\G_{t}$ acts non-trivially only on rows $t$ and $t-1$. As such, using the fact that $\ket{\psi_{t}} = \G_{t}\ket{\psi_{t-1}}$, the reduced state $\rho_{t}$ can be related to $\rho_{t-1}$ as,
\begin{align}
\rho_{t} &= \Tr_{\tau < t}\left[\G_{t} \left( \Xi_{t-1} \otimes \ketbra{\Omega_{t} }{\Omega_{t}} \right) \G^{\dag}_{t}\right]  ~, \\
&= \Tr_{t-1}\left[\G_{t} \rho_{t-1} \otimes \ketbra{\Omega_{t} }{\Omega_{t}} \G^{\dag}_{t}\right] ~.
\label{eqn:row_evo}
\end{align}
Here, $\ket{\Omega_{t}}$ is a $1D$ product state of all empty sites describing the row $t$ just before the update.

This clearly shows that the evolution of the $(1+1)D$ QCA induces a discrete time-evolution of a $1D$ quantum state $\rho_{t}$. While the evolution of the $2D$ state $\ket{\psi_{t}}$ is unitary, so that the state remains always pure, the evolution of such a $1D$ QCA is in general non-unitary and the state $\rho_{t}$ mixed. In this sense, the unitary $(1+1)D$ QCA can induce a non-unitary $1D$ QCA.

\subsection{Relation to system-environment evolution}
The evolution of the $1D$ state $\rho_{t}$ can be related to the usual evolution of a coupled system-environment under Markovian assumptions. For a system state $\rho_{S}$ and environment-state $\phi_{\mathcal{E}}$, the system state at time $t$ can be written as,
\begin{align}
    \rho_{S}(t) = \Tr_{\mathcal{E}}\left[ U_{t} \rho_{S} \otimes \phi_{\mathcal{E}} U^{\dagger}_{t}\right] ~,
\label{eqn:system_env}
\end{align}
where $U_{t}$ is the joint system-environment unitary.

Such an evolution can be made equivalent to the reduced evolution \eqref{eqn:row_evo} by choosing $U_{t} = S \G_{t}$ where $S\ket{n m} = \ket{m n}$ is a SWAP gate acting on the system and environment. In that case,
\begin{align}
    \rho_{S}(t) &= \Tr_{\mathcal{E}}\left[ S \G_{t} \rho_{S} \otimes \phi_{\mathcal{E}} \G^{\dagger}_{t} S^{\dagger}\right] ~, \\
&= \Tr_{S}\left[\G_{t} \rho_{S} \otimes \phi_{\mathcal{E}} \G^{\dagger}_{t}\right] ~.
\end{align}

Taking $\rho_{S} = \rho_{t-1}$ and $\phi_{\mathcal{E}} = \ketbra{\Omega_{t}}{\Omega_{t}}$ then reproduces the $1D$ state evolution of Eq. \eqref{eqn:row_evo}.

\subsection{Relation to unitary $1D$ QCA}

As a simple example of an evolution where the purity of $\rho_{t}$ is preserved, one can consider a local update gate of the form $G = U \otimes \mathds{1}$, with $U = \exp\left[- i \delta t (h_{1,2} + h_{2,3})\right]$, $\delta t = 0.01$ and $h_{i,j} = \sigma_{i}^{x}n_{j} + n_{i}\sigma^{x}_{j}$. This is the choice that we made to produce the plot of Fig.~\ref{fig:Seed_QCA_Schematic_a}(d). In this case we can write the global update gate as $\G_{t} = \mathcal{U}_{t-1} \otimes \mathds{1}_{t}$, where $\mathcal{U}_{t}$ is a unitary, ordered product of the $U$s (one per target site) which act on row $t-1$.

It is important to notice that this choice of $\G_{t}$ does not entangle the rows $t$ and $t-1$ (indeed in this example, the row $t$ is not modified at all). As such, if SWAP gates are applied following the global update, then the evolution of $\rho_{t}$ is unitary as shown by the following iterative equation:
\begin{align}
\rho_{t} &= \Tr_{t-1}\left[S\left(\mathcal{U}_{t} \rho_{t-1} \mathcal{U}^{\dagger} _{t} \otimes \ketbra{\Omega_{t}}{\Omega_{t}}\right)S^{\dagger}\right] ~, \\
&= \Tr_{t}\left[\left(\mathcal{U}_{t} \rho_{t-1} \mathcal{U}^{\dagger}_{t}\right)\otimes \ketbra{\Omega_{t}}{\Omega_{t}}\right] ~, \\
&= \mathcal{U}_{t} \rho_{t-1} \mathcal{U}^{\dagger}_{t} ~.
\end{align}

\section{Relations to Classical Model}

\subsection{Pure state representation of classical model}

The classical probability distribution of a PCA can be encoded in a so-called probability vector \cite{Harada2019} via the relation
\begin{align}
\ket{\psi^{\text{cl}}_{t}} = \sum_{m} P_{t}(m) \ket{m} ~,
\label{eqn:classical_pure_state}
\end{align}
where one has $\sum_mP_t(m)=1$. Here, indeed, $m$ labels the set of orthonormal states of definite occupation (i.e.~the classical configurations), while $P_{t}(m)$ is the probability of the state $m$ occurring in the PCA at time $t$.
The norm of this state provides the ``purity" of such a classical probability distribution, that we call $\gamma^{\rm cl}$, 
\begin{align}
\braket{\psi^{\text{cl}}_{t}|\psi^{\text{cl}}_{t}} &= \sum_{m} P_{t}(m)^{2} ~,
\\
&= \gamma^{\text{cl}} ~.
\end{align}
The value of $\gamma^{\rm cl}$ is $1$, denoting purity of the state, if and only if the probability is such that $P_t(m)=1$, only for a single state $\ket{m}$ and zero for all the others. 

Equivalently, the distribution $P_{t}(m)$ can be encoded in a diagonal density matrix as,
\begin{align}
    \rhocl&= \sum_{m} P_{t}(m)\ketbra{m}{m} ~.
\end{align}
This can be obtained by mapping $\ket{m} \to \ketbra{m}{m}$ in Eq. \eqref{eqn:classical_pure_state}. The purity of this density matrix is  equal to $\gamma^{\text{cl}}$. This classical purity can be used to define a (classical) Renyi-$2$ entropy for the probability distribution as  $S_{2}^{\text{cl}} = - \ln\left(\gamma^{\text{cl}}\right)$.

\subsection{Coherence and difference between entropies of the classical and of the quantum  models}

In a $(1+1)D$ classical model, $P_{t}(m)$ represents the probability distribution of $1D$ states of definite density (labelled by $m$) of the $t^{th}$ row at time $t$ \cite{Harada2019}. 
When extending such a PCA to a quantum model via $(1+1)D$ QCA, the $1D$ reduced density matrix of the quantum system, $\rho_{t}$, features, as the diagonal terms, the entries of $\rhocl$. The main difference between $\rho_t$ and $\rhocl$ lies in the presence of coherence terms in the density matrix $\rho_t$. As a consequence,  the purity of the states $\rho_{t}$ and $\rhocl$ differ by a contribution which is equal to the $\ell_2$-norm of the coherence, $C_{2} = \gamma - \gamma^{\text{cl}}$, where we have defined $\gamma=\Tr\left(\rho_t^2\right)$.

In terms of $C_{2}$, the difference between the quantum Renyi-$2$ entropy, $S_2=-\ln(\gamma)$, and the classical one $S_2^{\rm cl}$ is
\begin{align}
    \Delta &=  S_2 - S_{2}^{\text{cl}}, \\
    &= -\ln\left[ 1 + \frac{C_{2}}{\gamma^{\text{cl}}} \right] ~, \\
    &= -\ln\left[ 1 + e^{S_{2}^{\text{cl}}}C_{2} \right] ~.\\
\end{align}

At criticality, $S_{2}^{\text{cl}} \to 0$ as $t \to \infty$ and the absorbing state (which is pure) is approached. We can then expand $e^{S_{2}^{\text{cl}}}C_{2}$ as,
\begin{align}
    e^{S_{2}^{\text{cl}}}C_{2} = C_{2} + C_{2} S_{2}^{\text{cl}} + C_{2} S_{2}^{\text{cl}}/2 + ... ~.
\end{align}
Keeping only the leading order term and using that $C_{2} \ll 1$ when $t \to \infty$ near criticality, further expanding $\Delta$ gives, at first order in the coherence,
\begin{align}
    \Delta = - C_{2} + ... ~.
\end{align}

Therefore, we can approximate,
\begin{align}
    S_{2} (t) \approx S_{2}^{\text{cl}} (t) - C_{2} (t) ~,
\end{align}
for $t \gg 1$ and when the system is close to criticality. 

If, for a given initial condition, the coherence term $C_2(t)$ decays and approaches zero, we can expect $S_{2} \to S_{2}^{\text{cl}}$. In this case, critical exponents of the quantum and of the classical entropies would agree.

\section{Doubled-Space representation}

\subsection{States and Observables}

To simulate the evolution of $\rho_{t}$ with MPSs, we use a  doubled-space representation of the reduced density matrix. In this framework, operators are mapped to vectors according to the isomorphism $\ketbra{m}{n} \to \ket{m} \otimes \ket{n}$ \cite{Choi1975}. One thus has, 
\begin{align}
\rho(t)=\sum_{m,n}\rho_{mn}(t)\ketbra{m}{n}\to \ket{\rho(t)}=\sum_{mn}\rho_{mn}(t)\ket{m}\otimes \ket{n}\, .
\end{align}
Here, the symbol $\otimes$ indicates a product between the two parts of the doubled-space, rather than between rows of the QCA as in the main text and in the previous sections of this supplemental material.

Within such a representation, expectation values of observables can be calculated as
\begin{align}
O(t) &= \Tr\left[ \rho(t) \hat{O}\right] = \braket{\mathds{1} | \hat{O}_{L} | \rho(t)} ~ ,
\end{align}
where $\hat{O}_{L} = \hat{O} \otimes \mathds{1}$ and $\ket{\mathds{1}} = \sum_{m} \ket{m} \otimes \ket{m}$ is the doubled-space vector  representation of the identity operator. 

The purity of $\rho(t)$ can then be calculated as
\begin{align*}
\gamma &= \Tr\left[\rho(t)^{2}\right] = \sum_{m,n} \rho_{mn} \rho_{nm} ~ \\
&=  \sum_{m,n} \rho_{mn} \rho_{mn}^{*} \\
&= \braket{\rho(t)|\rho(t)} ~.
\end{align*}
We recall here that the vector representation of the state $\rho(t)$ is normalized in such a way that $\braket{\mathds{1}|\rho(t)}=1$, so that the purity is in general different from $1$. From $\gamma$, the second-order Renyi entropy can be calculated as $S_{2} = - \ln\left(\gamma\right)$.

\subsection{Time-evolution}
In the doubled-space representation, the evolution equation \eqref{eqn:row_evo} of the reduced $1D$ QCA can be expressed as the action of a linear map, $\Lambda_{t}$, onto the state $\ket{\rho_{t-1}}$, i.e., 
\begin{align}
\ket{\rho_{t}} = \Lambda_{t}\ket{\rho_{t-1}} ~.
\label{Lam-def}
\end{align}
In what follows, the map $\Lambda_{t}$ is expressed as a matrix product operator, which allows for the application of matrix product state methods to determine the evolution of the matrix product state representation of $\ket{\rho_{t}}$.

To find the form of the map $\Lambda_t$ we can proceed as follows. We first extend the vector representation of $\ket{\rho_{t-1}}$ in order to include also the row $t$ and its doubled-space component. These are both initialized with all particles in the down state. We thus have 
\begin{equation}
\ket{\tilde{\rho}_{t-1}}=\ket{\rho_{t-1}}\otimes \ket{\Omega_t}\otimes \ket{\Omega_t}=\sum_{m,n}\rho_{mn}(t-1)\ket{m_{t-1}}\otimes \ket{n_{t-1}}\otimes \ket{\Omega_t}\otimes \ket{\Omega_t}\, ,
\label{Double-repr-state}
\end{equation}
where the subscripts $t-1,t$ indicate to which row the vectors entering in the tensor product belong. The second and the fourth entries of the tensor product in the above equation, form the doubled-space components needed to vectorize the density matrix $\rho_{t-1}\otimes \ket{\Omega_t}\bra{\Omega_t}$. The update is obtained, in the density matrix formalism, by applying the global update gate $\mathcal{G}$ on both sides of the density matrix and tracing out the degrees of freedom on row $t-1$ [c.f.~Eq.~\eqref{eqn:row_evo}]. In the doubled-space representation this is achieved via 
$$
\ket{\rho_t}=\bra{\mathds{1}_{t-1}}\mathcal{G}_{13} \mathcal{G}_{24}^*\ket{\tilde{\rho}_{t-1}}\, .
$$
Here, $\ket{\mathds{1}_{t-1}}$ is the doubled-space representation of the identity operator with support solely on row $t-1$, and implements the partial trace over the associated space. The notation $(\cdot)^*$ indicates (element-wise) complex conjugation. Furthermore, we have defined $\mathcal{G}_{ij}$ to be the global update which takes, as control sites, the ones described by the vector in the $i$th entry of the tensor product of Eq.~\eqref{Double-repr-state} and, as target sites, those described by the vector in the $j$th entry of the tensor product.

This result shows that the map $\Lambda_t$ appearing in Eq.~\eqref{Lam-def}, acting directly on the vector $\ket{\rho_{t-1}}$, is given by the operator
\begin{align}
\Lambda_{t} = \bra{\mathds{1}_{t-1}}\G_{13} \G_{24}^{*} \ket{\Omega_{t}}\otimes \ket{\Omega_t} ~.
\label{eqn:lamda_def}
\end{align}

\section{Tensor Network Representation of Global Update in Doubled-Space}

In this section we present a method for representing the global update operator $\Lambda_t$ [\textit{c.f.} Eq. \eqref{eqn:lamda_def}] as a matrix product operator (MPO). When further representing the state $\ket{\rho_{t}}$ as a matrix product state (MPS), standard methods for MPSs can be applied to approximate the time-evolution $\ket{\rho_{t}} = \Lambda_{t}\ket{\rho_{t-1}}$. These can be found in, e.g., \cite{Schollwock2011,Montangero2018,Paeckel2019,Ran2020}. 

While this approach is limited to studying reduced dynamics on finite size lattices, it has the significant advantage that it can be integrated into existing tensor network algorithms for dynamics based on MPOs and MPSs, which are extremely common and have been highly optimised.

\subsection{MPO Representation of $\G$}

\begin{figure}[ht]
\centering
\includegraphics[width=1\linewidth]{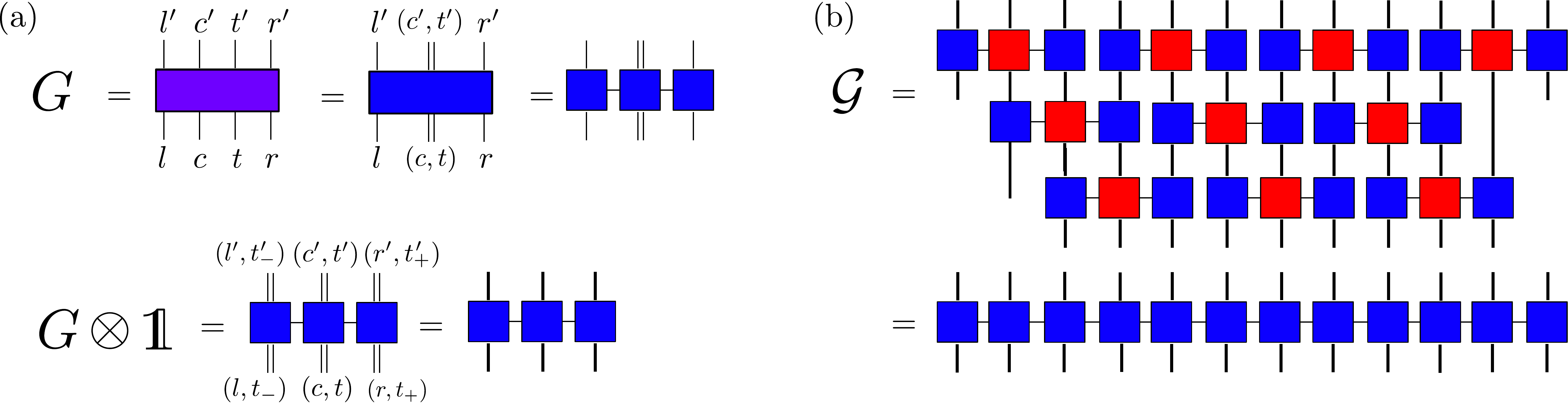}
\caption{\textbf{Construction of Global Update $\G$ as an MPO} (a) $G$ is represented as a three-site MPO. For a three-site neighbourhood $G$ is initially represented by a tensor of order-$8$. Indices relating to the control sites are labelled as $l,c,r$ (left/centre/right) while the target is labelled $t$. This is then reshaped into a tensor of order-$6$ by collecting the indices for the central control site and the target site as indicated in the figure. The resulting tensor can be decomposed via, e.g., singular value decomposition \cite{Schollwock2011}, to form a three-site MPO. To ensure all local dimensions are equal (corresponding to the vertical legs in the diagram) here we attach the identity matrix, $\mathds{1}$, onto the sites to the left and right of the target site in the same row (indicated by $t_{-}$ and $t_{+}$ respectively). (b) The MPO representation of $\G$ is then formed by applying $G$ across the system, once per target site. Here, for illustration, we have chosen a system of size $L =10$. The target site for each application of $G$ is highlighted in red. Note that the $\G$ obtained has support over $12$ sites, which includes a left and a right boundary site. In the main text, these are always assumed to be in the empty state. Since in general $G$ acting on different sites do not commute, the order in which they are applied to form $\G$ matters. Here, we have chosen a simple partitioned ordering that minimises the number of layers as shown, though in principle any can be chosen.}
\label{fig:MPO_Construction_Gt}
\end{figure}

Before turning to $\Lambda$, we construct an MPO representation for $\G$, via the procedure illustrated in Fig. \ref{fig:MPO_Construction_Gt}, which uses the common diagrammatic notation available for tensor networks \cite{Schollwock2011}.  We will consider here the case of three-site neighbourhoods, but this can be generalised. First, $G$ -- which is a four-body operator here acting on three control sites and a single target site -- is represented as a three-site MPO, see Fig. \ref{fig:MPO_Construction_Gt}(a).  To construct $\G$ from these, one simply chooses a particular ordering of target sites and contracts the MPO representations of $G$ for each target site together, according to this ordering, see Fig.~\ref{fig:MPO_Construction_Gt}(b).

\subsection{MPO Representation of $\Lambda$}

\begin{figure}[ht]
\centering
\includegraphics[width=1\linewidth]{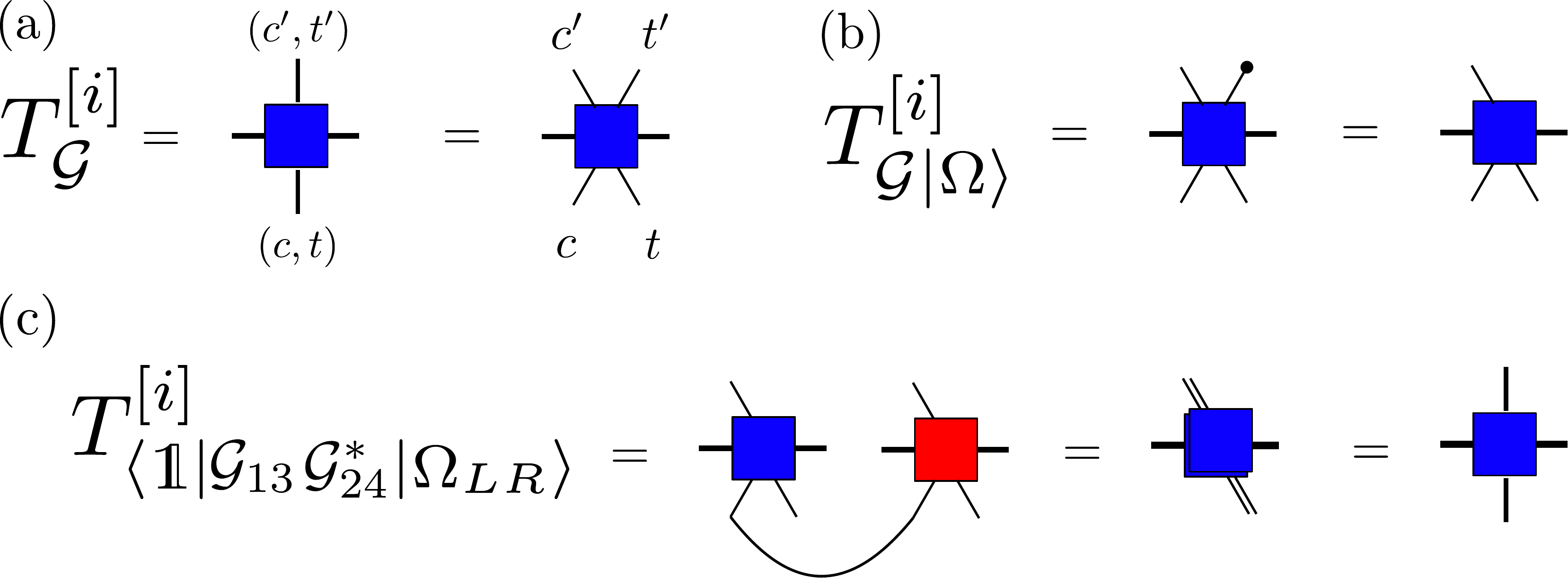}
\caption{\textbf{Construction of $i^{th}$ tensor in the MPO Representation of $\Lambda$.} (a) The physical indices of the $i^{th}$ tensor in the MPO representation of $\G$ (see Fig. \ref{fig:MPO_Construction_Gt}(b)) are factorised. Here, $c$ labels an index belonging to the control row, while $t$ labels an index belonging to the target row. This results in an order-$6$ tensor. (b) Since the target row is always initialised as a product state of all empty sites, the vector $\ket{\circ}$ -- indicated by a black circle in the figure -- can be applied to the appropriate index at this stage. The resulting tensor, which is of order-$5$, occurs in the MPO representation of $\G\ket{\Omega}$. (c) The tensors for the MPO representation of $\Lambda$ are formed by taking the (elementwise) complex conjugate of a copy of $T^{[i]}_{\G\ket{\Omega}}$ (indicated here in red). A tensor product with the previous tensor would provide the $i^{th}$ tensor in the MPO representation of $\G_{13}\G^{*}_{24}\ket{\Omega_{LR}}$ where $\ket{\Omega_{LR}} = \ket{\Omega} \otimes \ket{\Omega}$ is the doubled-space representation of $\ketbra{\Omega}{\Omega}$. Contracting the two indices indicated in the control row ($t-1$) then gives the $i^{th}$ tensor in the MPO representation of $\Lambda$ [\textit{c.f.} \eqref{eqn:lamda_def}].}
\label{fig:MPO_Construction_DS}
\end{figure}

To evolve $\rho_{t}$ represented in the doubled-space via MPS, we need to find a representation of $\Lambda_{t}$ [\textit{c.f.} \eqref{eqn:lamda_def}] as an MPO in this space. With an MPO representation of $\G$, this can be achieved by repeating the same general procedure for each tensor individually, as illustrated in Fig. \ref{fig:MPO_Construction_DS}. 

Labelling the $i^{th}$ tensor in a given MPO representation as $T^{[i]}$, we begin by taking the $i^{th}$ tensor of the MPO representation of $\G$, which we then denote as $T^{i}_{\G}$, and factorising the physical indices, see Fig. \ref{fig:MPO_Construction_DS}(a). The tensors representing $\G\ket{\Omega}$ can then be obtained from this by applying a vector representing $\ket{\circ}$ to the appropriate leg, see Fig. \ref{fig:MPO_Construction_DS}(b).

The tensors of $\Lambda$ can then be obtained in two stages. First, a copy of the current tensor is made and the (elementwise) complex conjugate is taken. This new tensor appears in the MPO representation of $\G^{*}\ket{\Omega}$. This is combined with the previous tensor by contracting over the legs shown in Fig. \ref{fig:MPO_Construction_DS}(c), which correspond to those traced-out via the application of $\ket{\mathds{1}_{t-1}}$ in the doubled space. The indices of the resulting tensor (which is of order-$8$) are then collected together as shown to form an order-$4$ tensor. This is the $i^{th}$ tensor in the MPO representation of $\Lambda_{t}$. Repeating this procedure for all $i = 1,2,...,L+2$ (inclusive of the two boundaries) then produces the desired MPO representation. 

To evolve a state $\ket{\rho_{t-1}}$ defined as an MPS over $L$ sites with this MPO in the main text, we first expand $\ket{\rho_{t-1}}$ with an empty site to the left and right resulting in a state with support on $L+2$ sites. The MPO for $\Lambda$ is then applied variationally (see e.g. \cite{Paeckel2019}) before tracing out the left-most and right-most sites to form an approximation of $\ket{\rho_{t}}$.

\section{Errors in Estimation of Critical Exponents}

In this section we discuss the errors for the estimates of the critical exponents of the models. These values are collected in Table \ref{table:error_estimates}. For each exponent, the error presented in the main text is  taken as the largest of the estimated errors from the various sources that we describe below.

\subsection{QDKCA model}

In determining the critical exponents, one needs to consider errors associated to two primary (though not independent) sources. The first are those associated with  estimating the location of the critical point. Such an error is difficult to quantify and can, potentially, be very important. In this regard, however, the QDKCA model holds a key advantage over the QBBR, given that the associated classical DKCA has been studied extensively and the critical point determined to a very high precision. As such, although it remains difficult to quantify the relevance of this error, we can expect this to be negligible, for the QDKCA, as compared to other sources. This is the assumption that we make, and we completely neglect this source of error for the QDKCA.

The remaining error sources for the QDKCA concern the accurate approximation of $S_{2}$ and $C_{2}$, and any associated effective exponents. For fixed $L, p_{1}, p_{2}$ and up to a given $t$, this error is controlled by the bond-dimension of the MPS, $\chi$. As $\chi \to \infty$ this error must vanish. We observe that the tendency of finite-$\chi$ errors are to produce an artificial curvature upwards in $S_{2}(t)$. This can be seen most clearly in the data of Fig. \ref{fig:Results}(c) for $p_{1} = 0.65$ between $t = [100,200]$ where, rather than tending to a stationary value, the curve bends upwards. As this curve is not used in approximating the critical exponents, this does not directly affect estimates. However, to avoid under-estimation of errors, the tendency of these finite-$\chi$ effects does alter our error analysis indirectly for the QBBR model where finite-$\chi$ effects are larger and the critical point is not precisely known. See below for a discussion of this issue.

The value of $\chi$ required for an acceptable error can depend strongly on the model's parameters. To estimate the error due to finite-$\chi$, we recalculate each quantity of interest -- in this case the critical exponents -- using approximations with a $\chi$ being half of the one used for the estimate.

The second source of error that concerns the accurate approximation of $S_{2}$ and $C_{2}$ -- which we would like to obtain free of finite-size effects  -- is that of finite-$L$. To estimate the relevance of this, we follow a similar procedure as for $\chi$ and recalculate quantities using simulations with $L/2$, but otherwise fixed parameters.

Finally, an additional issue when estimating the values of critical exponents are finite-time errors. To quantify these, we recalculate the estimated quantities (via fits or averages) but over an interval of time that is half of that used for the original estimate, starting at the same initial time.

As discussed in the main text, $\qe$ is estimated via a power-law fit of the form $ t^{-\qe}$ to $S_{2}(t)$ approximated with $\chi = 128, L = 256$ between $t = \left[50,200\right]$ for $p_{1} = 0.645$. This gives $\qe = 0.21265$ (all estimated quantities are given in this section to five decimal places while estimated errors are given to three significant digits). Repeating this procedure with $S_{2}(t)$ calculated using $\chi = 128, L = 128$ gives $\qe = 0.21270$. With $\chi = 64, L = 256$ we obtain $\qe = 0.21263$. Finally, with $\chi = 128, L = 256$ between $t = \left[50,125\right]$ gives $\qe = 0.21635$. Taking the absolute difference between each of these estimates and the original, we approximate the errors due to each source to be: $5.37 \times 10^{-5}$, $2.21 \times 10^{-5}$,  and  $3.70 \times 10^{-3}$ for finite-$L$, finite-$\chi$ and finite-time respectively.

To estimate errors in $\qc$ the same procedure is used as with $\qe$ for error due to finite-$\chi$ and finite-time. In this case, the original estimate is obtained by averaging over the effective exponent $\qc(t)$ from $t= [450,550]$ to obtain $\qc = 2.96486$. Repeating the error calculations as for $\qe$ obtains: $2.88 \times 10^{-3}$ for finite-$\chi$ errors and $3.57 \times 10^{-3}$ for finite-time errors (this being obtained by averaging over $t = [450,500]$). Due to the long times required, estimating errors due to finite-$L$ from calculations with $L=128$ leads to a significant overestimation, as for this time the state is already approaching the absorbing state. As such, to estimate finite-$L$ errors in this case we instead compare the results with those obtained from a method, free of finite-size effects, which provides an estimated error of $1.3 \times 10^{-3}$. Details of this method can be found in \cite{Gillman2021}. These data were produced using $\chi = 64$ but otherwise equal model parameters.

\subsection{QBBR model}

While the error due to the determination of the critical point was assumed to be minimal for the QDKCA, in the QBBR model -- since the classical BBR model is much less studied -- this can play an important role. This is particularly true since the estimation of critical exponents is very sensitive to the distance from the true critical point.

For APTs, a standard procedure for estimating errors in this regard is to bound the value of a critical exponent using curves that are known to be in a given phase \cite{Henkel2008}. In the case of a power-law decay at a critical point, curves in the inactive phase (which decay more rapidly than the critical curve) can provide upper-bounds to the exponent. Similarly, curves in the active phase can provide lower-bounds. By simulating a set of curves on some grid up to a given time, one then identifies curves in a given phase and uses these for the error estimate. To decrease the error in such an estimate, a finer grid can be used. However, distinguishing the phase of a given curve will require longer and longer times the closer it is to the critical point. As such, ultimately, the limitation in this procedure stems from the maximum time that can be accurately approximated.

For the QBBR model, we estimate errors due to the uncertainty on the critical point using a simple grid search. That is, from the values of $p_{1}$ chosen -- which form the ``grid" -- we select as the estimated critical point the one that results in a curve $S_{2}(t)$ that best approximates a power-law. Errors are then estimated by recalculating quantities using two other curves, one from the inactive phase, and one from the active phase. For the curves of $S_{2}(t)$ presented in the main text [\textit{c.f.} Fig. \ref{fig:Results}(c)] the one with $p_{1} = 0.61$ best approximates a power-law and is therefore chosen as the estimated critical point. The curve below this one,  with $p_{1} = 0.6$, shows clear behaviour characteristic of the absorbing (inactive) phase. As such, we take this curve to provide an upper bound on $\qe$. To provide a lower-bound, the curve with $p_{1} = 0.62$ displays curvature indicative of belonging to the active phase. However, as mentioned previously, finite-$\chi$ errors tend to produce this an upward curvature artificially. As such, to avoid underestimation of this error, we instead take the curve with $p_{1} = 0.625$ in this instance.

Taking the estimated critical point, $p_{1} = 0.61$, finite-$L$, finite-$\chi$, and finite-time errors are estimated for $\qe$ as for the QDKCA. The original estimate, made with $\chi = 128, L = 256$ over $t = [50,200]$ gives $\qe = 0.26716$. The estimated errors for finite-$L$, finite-$\chi$ and finite-time are  $4.10 \times 10^{-4}, 1.09 \times 10^{-2}$ and $3.75 \times 10^{-3}$ respectively. To estimate the errors due to the uncertainty in the critical point, we perform two power-law fits. The first, for $p_{1} = 0.60$ with $\chi = 128, L = 256$ over $t = [10,40]$ (where the curve displays an approximate power-law) gives $\qe = 0.32250$ leading to an estimated error (via the absolute difference) of $5.53 \times 10^{-2}$. The second with $p_{1} = 0.625, \chi = 128, L = 256$ over $t = [10,40]$ gives $\qe = 0.20171$ and an error of $6.54 \times 10^{-2}$. 

The estimates of $\qc$ for the QBBR model proceed in a similar fashion. Taking the same $p_{1} = 0.61$ to estimate the critical point and averaging over $t=[150,200]$ gives $\qc = 2.73746$. The finite-$L$,  finite-$\chi$ and finite-time errors (calculated by averaging over $t = [150,176]$) are estimated as $7.11 \times 10^{-3}, 1.56 \times 10^{-3}$ and $3.75 \times 10^{-3}$ respectively.

For estimating the errors related to determination of the critical point, only values of $p_{1}$ for which $\qc(t)$ is approximately constant over a substantial interval of $t$ can be used. In practice, we find that this excludes values of $p_{1}$ from the active phase as the effective exponent tends to diverge, meaning no such interval can be found. As such, here we construct a simple estimate using the curve that previously provided the lower bound for $\qe$ (i.e. the curve with $p_{1} = 0.60$) and assume a symmetric error about the estimate of $\qc$. Calculating $\qc$ just as for $p_{1} = 0.61$ but with $p_{1} = 0.60$ provides $\qc = 2.47247$, giving an associated error of $2.65 \times 10^{-1}$. 

The error estimates from the various sources discussed are summarised in Table \ref{table:error_estimates}.

\begin{table}
  \centering
  \begin{tabular}{|c|c|c|c|c|c|c|}
  \hline
   & Estimate & Finite-$L$ & Finite-$\chi$& Finite-time & Critical Point (Lower Bound) & Critical Point (Upper Bound) \\
    \hline
    QDKCA : $\qe$ & $0.21265 \pm 0.00370$ &  $5.37 \times 10^{-5}$ & $2.21 \times 10^{-5}$ & $3.70 \times 10^{-3}$ & $-$ & $-$\\
    \hline
    QDKCA: $\qc$ & $2.96486 \pm 0.00357$ &  $1.3 \times 10^{-3}$ & $2.88\times 10^{-3}$ & $3.57 \times 10^{-3}$ & $-$ & $-$\\
    \hline
    QBBR: $\qe$ & $0.26716 \pm 0.0654$ &  $4.10 \times 10^{-4}$ & $1.09 \times 10^{-2}$ & $3.75 \times 10^{-3}$ & $6.54 \times 10^{-2}$ & $5.53 \times 10^{-2}$\\
    \hline 
    QBBR: $\qc$ & $2.73746 \pm 0.265$ &  $7.11 \times 10^{-3}$ & $1.56 \times 10^{-3}$ & $3.75 \times 10^{-3}$ & $2.65 \times 10^{-1}$ & $2.65 \times 10^{-1}$\\
    \hline
  \end{tabular}
  \caption{\textbf{Error Sources and Estimates for Critical Exponents} The estimates of the critical exponents examined in the main text, along with error estimates due to various sources. Details on the calculation of each estimate are contained in the text of the supplemental materials.} \label{table:error_estimates}
\end{table}

\end{document}